\documentclass[twocolumn]{aastex61}
\usepackage{graphicx}				
\usepackage{xcolor}
\usepackage[sort&compress]{natbib}
\usepackage[hang,flushmargin]{footmisc}
\usepackage[counterclockwise]{rotating}
\usepackage{amsmath,natbib,graphicx}
\usepackage{lipsum}
\usepackage{gensymb}

\shortauthors{Checlair et al.}
\usepackage{wrapfig}
\shorttitle{Tidally locked ocean diffusion}
\usepackage{enumitem,xcolor}

\begin{document}
\graphicspath{ {./} }
\DeclareGraphicsExtensions{.pdf,.eps,.png}

\title{No snowball on habitable tidally locked planets with a dynamic ocean}
\accepted{at ApJL September 25 2019}

\author{Jade H. Checlair}
\affiliation{Department of the Geophysical Sciences, University of
  Chicago, 5734 South Ellis Avenue, Chicago, IL 60637}

\author{Stephanie L. Olson}
\affiliation{Department of the Geophysical Sciences, University of
  Chicago, 5734 South Ellis Avenue, Chicago, IL 60637}
  
\author{Malte F. Jansen} 
\affiliation{Department of the Geophysical Sciences, University of
  Chicago, 5734 South Ellis Avenue, Chicago, IL 60637}
  
\author{Dorian S. Abbot}
\affiliation{Department of the Geophysical Sciences, University of
  Chicago, 5734 South Ellis Avenue, Chicago, IL 60637}
 
\correspondingauthor{Jade Checlair}
\email{jadecheclair@uchicago.edu}


\begin{abstract}
 
Terrestrial planets orbiting within the habitable zones of M-stars are likely to become tidally locked in a 1:1 spin:orbit configuration and are prime targets for future characterization efforts. An issue of importance for the potential habitability of terrestrial planets is whether they could experience snowball events (periods of global glaciation). Previous work using an intermediate complexity atmospheric Global Climate Model (GCM) with no ocean heat transport suggested that tidally locked planets would smoothly transition to a snowball, in contrast with Earth, which has bifurcations and hysteresis in climate state associated with global glaciation. In this paper, we use a coupled ocean-atmosphere GCM (ROCKE-3D) to model tidally locked planets with no continents. We chose this configuration in order to consider a case that we expect to have high ocean heat transport. We show that including ocean heat transport does not reintroduce the snowball bifurcation. An implication of this result is that a tidally locked planet in the habitable zone is unlikely to be found in a snowball state for a geologically significant period of time. 

\end{abstract}
\section{Introduction}
A number of planets have been found orbiting in the habitable zones of nearby M-stars (TRAPPIST-1e, Proxima Centauri b, and LHS 1140b, \citet{anglada2016terrestrial,gillon2017seven,dittmann2017temperate}). These planets are prime targets for future characterization efforts due to high planet-to-star flux ratios. M-stars are dimmer than G-stars, so that their habitable zones are close enough that planets orbiting within them are likely to become tidally locked in a synchronously rotating 1:1 spin:orbit configuration (hereafter ``tidally locked'') \citep{kasting1993habitable}. The large occurence rates of these habitable zone planets orbiting M-stars motivates investigations into the habitability of tidally locked planets \citep[e.g.,][]{Joshi:1997,Segura:2005,Merlis:2010,Kite:2011,Wordsworth:2011p3221,Pierrehumbert:2011p3287,yang2013,leconte20133d,menou2013water,hu2014role,yang2014b,shields2016habitability,kopparapu2016inner,turbet2016habitability,barnes2016habitability,wolf2017assessing,del2017habitable,bolmont2017water,meadows2018habitability,abbot2018decrease,chen2018biosignature,way2018climates,yang2019ocean,jansen2019climates}.  

Terrestrial planets in the habitable zone of G-stars (such as Earth) are subject to periods of global glaciations, which are called snowball events. In particular, Earth is believed to have gone through a few Snowball events in its history \citep{Kirschvink92,Hoffman98, hoffman2017snowball}. The effect of snowball events on life is uncertain, but global glaciations on Earth are correlated with increases in atmospheric oxygen and in the complexity of life \citep{Kirschvink92,Hoffman98, Hoffman02,laakso2014regulation,laakso2017}. On the other hand, global glaciations could be problematic for any pre-existing complex life. 

These snowball events are a result of the existence of climate bifurcations, bistability, and hysteresis in rapidly rotating planets. Bistability means that two different climate states (a ``Warm'' ice-free state and a ``Snowball'' ice-covered state) can exist for the same external forcing (CO$_2$, stellar flux, etc.). This is a result of a basic nonlinearity called the ice-albedo feedback caused by the difference in top-of-atmosphere albedo of ice/snow v.s. water \citep{Budyko-1969:effect,Sellers-1969:ae}. A planet in its Warm state may ``jump'' into a Snowball state when the stellar flux is decreased only slightly at a particular threshold. This is an example of a saddle-node bifurcation \citep{strogatz1994nonlinear}, called the ``snowball bifurcation'' in this context. This bifurcation (jump into a Snowball state) occurs at a critical ``glaciation'' stellar flux, but to deglaciate the planet the stellar flux must be increased to a much greater critical ``deglaciation'' value (another bifurcation). If the stellar flux is increased and decreased so that these bifurcations are crossed, its current state will be dependent on the path it has taken, which is referred to as climate hysteresis.

Recently, \citet{checlair2017no} argued using simple and complex atmospheric models that tidally locked planets are much less likely to exhibit the snowball bifurcation as a result of the strong increase in stellar irradiation as the substellar point is approached. Moreover, an M-star spectrum reduces the region of climate bistability by lessening the contrast in albedo between ice and ocean, making a snowball bifurcation less likely \citep{joshi2012suppression,shields2013effect,shields2014spectrum}.

Planetary heat transport, if efficient enough, could destabilize partially glaciated states on a tidally locked planet and reintroduce a snowball bifurcation as was discussed by \citet{checlair2017no}. However, the authors did not explicitly model ocean heat transport, motivating further investigation into its role in possibly retrieving the snowball bifurcation. Considering dynamic oceanic processes could increase the overall planetary heat transport, potentially making a bifurcation possible. \citet{hu2014role} previously modeled a tidally locked planet orbiting an M-star using an atmospheric GCM (CAM3) coupled to a dynamic ocean. They did not directly look for bifurcations or hysteresis, but found a smooth transition from a partially glaciated to a fully ice-covered planet, suggesting the potential lack of a bifurcation. In this paper, we investigate the effects of including a dynamic ocean on the snowball bifurcation for tidally locked planets in a coupled ocean-atmosphere GCM (ROCKE-3D). 

\section{Methods}
\label{sec:methods}

\subsection{Model Description}
\label{sec:ModelDescription}

We performed our calculations using ROCKE-3D \citep{way2017resolving}, a fully coupled ocean-atmosphere general circulation model (GCM) that is derived from NASA Goddard Institute for Space Studies (GISS) ModelE2. ROCKE-3D has previously been used to examine the climate states of slowly rotating and tidally locked exoplanets, including warm and snowball climates \citep{del2017habitable,jansen2019climates,way2018climates}. The reader is referred to \citet{way2017resolving} and references therein for details regarding ROCKE-3D and its parent model, but we note that ROCKE-3D includes a thermodynamic-advective sea ice model. In contrast with previous studies of snowball bifurcations on tidally locked planets \citep{checlair2017no}, ROCKE-3D considers ice advection, allows for partial ice coverage within cells, and includes dynamical ocean heat transport. Our present investigation using ROCKE-3D thus represents a significant advance over prior work.

\subsection{Model Configurations}
\label{sec:ModelConfigurations}

Our experiments consider an Earth-like planet with regard to mass, radius, surface gravity, and atmospheric mass and composition (although we exclude O$_3$). Our simulated planet nonetheless differs from the Earth in several ways. We simulate a tidally locked planet in a circular orbit with a 50 day period around an M-star (Kepler 1649,  M5V). We chose the stellar spectrum from ROCKE-3D's default spectra. It is based on the BT-Settl spectrum with T$_{eff} = 3200$K, logg = 5, Fe/H = 0 \citep{allard2012models}. The surface of the planet is covered exclusively by ocean. The ocean in our model consists of 5 layers and is considerably shallower than Earth's, reaching a maximum depth of 189 m. In this configuration, running the model on 44 CPU cores requires on average 180 hours of wall time. A deeper ocean would increase the computational cost considerably. Our model configuration is summarized in Table~\ref{tab:params}.

\subsection{Modeled Scenarios}
\label{sec:ModeledScenarios}

\begin{figure*}[h]
\begin{center}
  \includegraphics[width=6.3in]{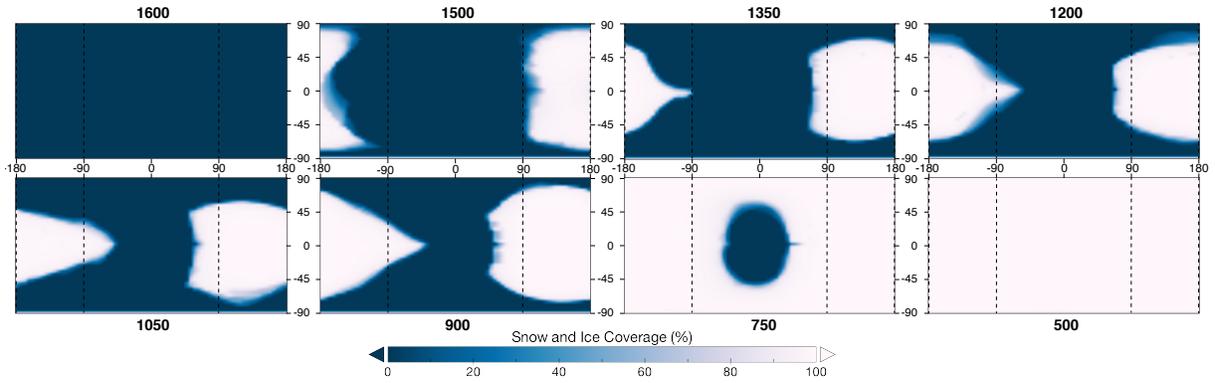}
 \end{center}
  \caption{Asymmetric progression of ice coverage on a tidally locked aquaplanet with a dynamic ocean (warm start). Steady-state snow and ice coverage (\% area) for decreasing stellar irradiation, from 1600 Wm$^{-2}$ (upper left) to 500 Wm$^{-2}$ (lower right) as a function of longitude (horizontal axis) and latitude (vertical axis). Although ice coverage monotonically increases with decreasing stellar irradiation, ice does not uniformly advance. Each panel is centered on the substellar point.}
  \label{fig:ice}
\end{figure*}

We first performed simulations at stellar irradiations, S, of 1600 Wm$^{-2}$ and 500 Wm$^{-2}$, yielding an equilibrated ice-free climate state and an equilibrated globally glaciated state, respectively. We then investigated potential climate hysteresis by using the ice-free equilibrated state (S=1600 Wm$^{-2}$) as the initial condition for our ``Warm Start'' experiments and the globally glaciated equilibrated state (S=500 Wm$^{-2}$) as the initial condition for our ``Cold Start'' experiments. For our Warm Start experiments, we restarted the model from an ice-free initial condition at progressively lower stellar irradiation. For our Cold Start experiments, we restarted the model from a globally-glaciated condition at progressively higher stellar irradiation. The Warm Start and Cold Start scenarios are shown at the upper left and lower right corners of Figures~\ref{fig:ice} and~\ref{fig:SST}. For both sets of experiments, we examined the resulting climate state at stellar irradiations of 1500, 1350, 1200, 1050, 900, and 750 Wm$^{-2}$. We diagnosed steady-state for each model scenario based on a net planetary radiation balance of +/- 0.2 Wm$^{-2}$. This was typically achieved on a timescale of $\sim$ 500 - 1000 model years.

\begin{table}
  \caption{\textbf{Model configuration}}
\label{tab:params}
\centering
\begin{tabular}{l|ll}
  Planetary Parameter  &  Value  \\
\hline
Mass & M$_{\earth}$ \\ 
Radius & R$_{\earth}$ \\ 
Surface pressure & 1 bar  \\
Atmospheric composition & Pre-industrial Earth (78$\%$ N$_2$, \\ 
 & 21$\%$ O$_2$, 1\% Ar, 285 ppm CO$_2$, \\
 & variable H$_2$O)\\
Continental configuration & Aquaplanet  \\
Ocean depth & 189 m (5 layers) \\
Obliquity & 0  \\
Eccentricity & 0 \\
Orbital period & 50 days \\
Rotation period & 50 days \\
Stellar spectrum & Kepler 1649 (M5V star) \\
Stellar irradiation & variable (500 - 1600 Wm$^{-2}$)
\end{tabular}
\end{table}

\section{Results}

\begin{figure*}[h]
\begin{center}
  \includegraphics[width=6.3in]{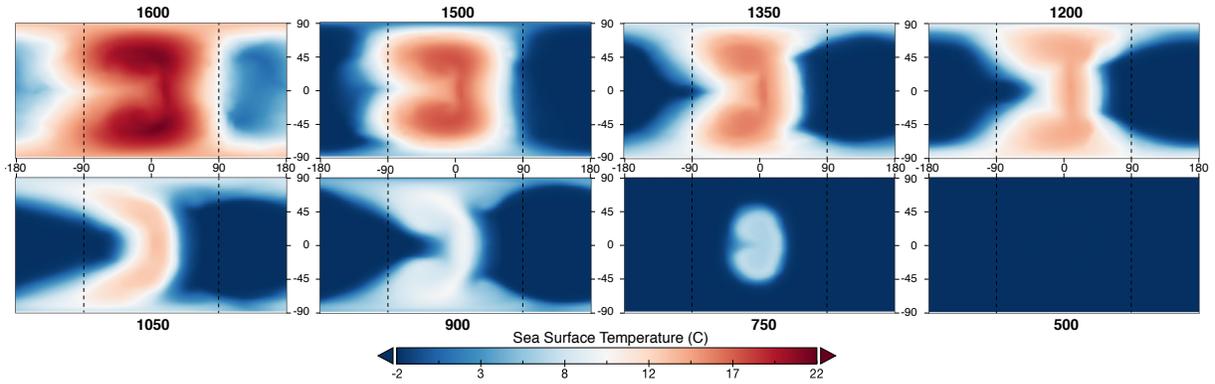}
 \end{center}
  \caption{Asymmetric progression of sea surface temperature on a tidally locked aquaplanet with a dynamic ocean (warm start). Steady-state sea surface temperature (\degree C) for decreasing stellar irradiation, from 1600 Wm$^{-2}$ (upper left) to 500 Wm$^{-2}$ (lower right) as a function of longitude (horizontal axis) and latitude (vertical axis). Ocean circulation redistributes heat poleward on the day side (Figure 3), promoting equatorial ice advance from the night side. Each panel is centered on the substellar point.}
  \label{fig:SST}
\end{figure*}

\begin{figure*}[h]
\begin{center}
  \includegraphics[width=6in]{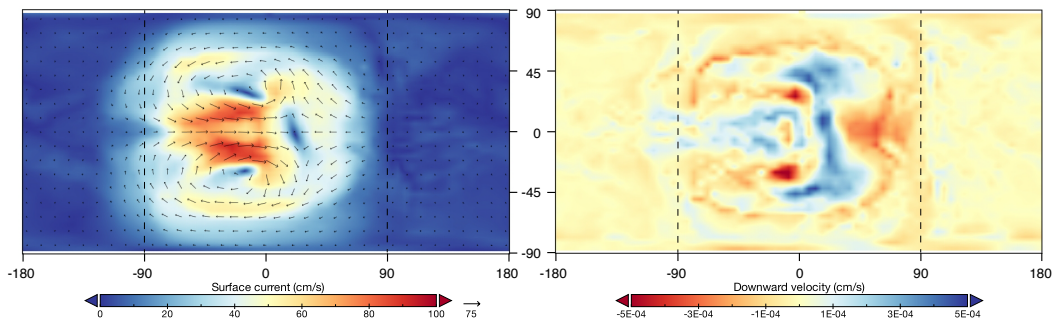}
 \end{center}
  \caption{Ocean circulation patterns that drive the asymmetric progression of ice coverage and surface temperature on a tidally locked aquaplanet with a dynamic ocean. Steady-state surface ocean currents (left) and vertical velocities (right) for the ice-free 1600 Wm$^{-2}$ scenario as a function of longitude (horizontal axis) and latitude (vertical axis). The plots are centered on the substellar point. The development of gyres (i.e. horizontally circulating currents) to the north-west and south-west of the substellar point preferentially warms the poles and limits eastward equatorial heat transport to the night side.}
  \label{fig:circu}
\end{figure*}

 The net effect of a dynamic ocean is to transport heat from the substellar region where energy is gained to the nightside where energy is lost. In addition to this, the ocean transports heat north and south of the substellar point, warming higher latitudes. To illustrate this, we first present maps of sea ice cover (Figure~\ref{fig:ice}), surface temperature (Figure~\ref{fig:SST}), and ocean surface currents (Figure~\ref{fig:circu}) for equilibrated annual-mean climates of the simulated tidally locked aquaplanet. As the stellar irradiation decreases, sea ice begins to form on the antistellar point and gradually progresses towards the substellar point in an asymmetrical manner (Figure~\ref{fig:ice}). Similarly, the increase in sea surface temperature is not radially uniform approaching the substellar point (Figure ~\ref{fig:SST}) as occurs for tidally locked planets without a dynamic ocean \citep[``Eyeball states,''][]{Pierrehumbert:2011p3287, Menou2015}. There are two areas north and south of the substellar point, and one smaller area eastward of it, that are warmer than the substellar point itself. These patterns are reflected in the sea ice cover patterns (Figure~\ref{fig:ice}) and are caused by circulation gyres, which can be seen in maps of upper layer oceanic circulation (Figure~\ref{fig:circu}). They were similarly found by \citet{hu2014role} and \citet{yang2019ocean}, who attributed them to superrotating equatorial surface winds caused by Rossby and Kelvin atmospheric and oceanic waves that drag sea ice from the night side to the dayside. This limits the overall heat transport to the nightside and shapes the asymmetrical progression of sea ice and sea surface temperature as stellar irradiation is varied.

\begin{figure*}[h!]
\begin{center}
  \includegraphics[width=7in]{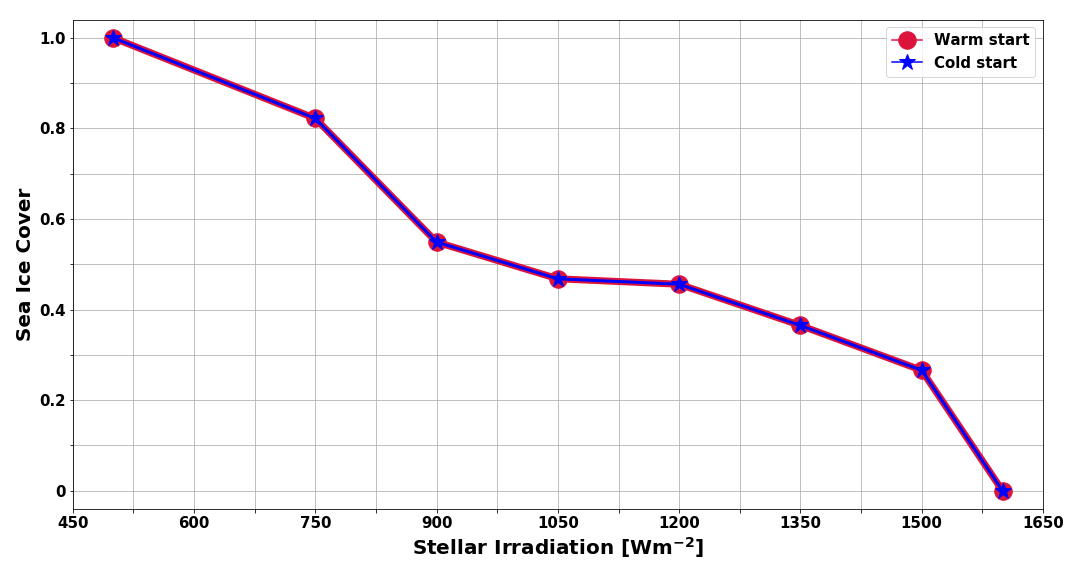}
 \end{center}
  \caption{No snowball bifurcation for the tidally locked planet. Global-mean equilibrated sea ice cover as a function of stellar irradiation. Blue stars correspond to a
  Cold Start (ice-covered planet) initialization and red circles to a
  Warm Start (ice-free planet) initialization. For all values of irradiation the planet equilibrates to the same final state regardless of its initial state.}
  \label{fig:rocke_bifurcation}
\end{figure*}

We next explore the possibility of bifurcations and bistability for a tidally locked planet with a dynamic ocean (Figure~\ref{fig:rocke_bifurcation}). As the stellar flux decreases, the planet's ice cover increases gradually, until reaching a globally ice-covered state between 500 and 750 Wm$^{-2}$. Similarly, if we increase the stellar flux, the planet gradually deglaciates and reaches an ice-free state between 1500 and 1600 Wm$^{-2}$. There is no bistability or hysteresis in the climate, as both the Warm Start and Cold Start configurations result in the same equilibrated climate at all stellar fluxes.

\section{Discussion}

An important issue in planetary habitability is the possibility for terrestrial planets to go through global glaciations, called Snowball events. Earth itself is believed to have gone through a number of snowball events in its history \citep{Kirschvink92,Hoffman98, hoffman2017snowball}. Although extreme glaciation might be expected to challenge life, snowball events are associated with increases in biological and ecological complexity in Earth's history. The reasons for this relationship are poorly understood, but may arise from accelerated evolution in the wake of environmental perturbation or enhanced nutrient delivery from glacial weathering and resulting increases in environmental oxygen following more than a billion years of apparent biogeochemical stasis \citep{Hoffman98, planavsky2010evolution, brocks2017rise, laakso2017theory}. Similar events may be required to trigger increasing complexity on other inhabited planets as well.

Snowball events are the result of the existence of the snowball bifurcation in the climate of rapidly rotating planets like the Earth. \citet{checlair2017no} previously showed using an intermediate-complexity GCM with a slab ocean that tidally locked planets such as those orbiting in the habitable zones of M-stars are unlikely to go through snowball bifurcations. This is a result of the strong increase in stellar irradiation as the substellar point is approached. However, they found that if planetary heat transport is increased to a critical value, the bifurcation may be recovered. Although this could theoretically reintroduce the snowball bifurcation, here we found that this does not occur in a complex ocean-atmosphere GCM with a dynamic ocean (ROCKE-3D). Instead, sea ice extent on tidally locked planets gradually increases as irradiation is decreased, with a wide range of states of intermediate glaciation (``Eyeball'' states).

The most important consequence of this work is that habitable tidally locked planets should not be found in a snowball state for an extended period of time. Consider a partially glaciated tidally locked planet with open ocean at the substellar point. If some perturbation such as a volcanic eruption or asteroid impact drove it into a snowball state, it would quickly return to the original, partially glaciated state. The reason is that tidally locked planets lack snowball bifurcations and bistability, so the partially glaciated state is the only stable state at the that stellar irradiation.

One caveat of our study is that the ocean depth is only 189 m. Increasing this depth could alter our results by increasing the ocean heat capacity and allowing deep circulation. In Earth’s ocean, the heat transport by the wind-driven gyres is concentrated in the upper ocean but can reach to around 500 m depth \citep[e.g.,][]{boccaletti2005vertical}. This suggests that a deeper ocean might lead to somewhat stronger ocean heat transport, possibly amplifying the effect of the gyres on sea ice cover. A much deeper ocean would also raise the possibility of a deep ocean overturning circulation between the day and night sides, which could additionally contribute to heat transport and potentially keep the night side ice-free at lower stellar irradiations. However, the strength of the deep ocean overturning is controlled by small-scale turbulent mixing, which needs to be parameterized in the GCM, and which is very hard to constrain \citep{wunsch2004vertical}. Using a similar GCM as presented here, \citet{yang2019ocean} found that the climate state of tidally locked aquaplanets is sensitive to ocean depth, although the relationship between ocean depth and global climate is not straightforward. In summary, the sensitivity of planetary climate and of the snowball bifurcation to ocean depth is complex and represents an intriguing topic for future work.

Land configuration may also affect ocean heat transport and glaciation dynamics. \citet{yang2019ocean} found that adding continental barriers reduces the heat transport from the day side to the night side on a tidally locked planet. Therefore, we chose to model a tidally locked aquaplanet to maximize ocean heat transport. Further work could be done to explore the effect that different land configurations have on planetary heat transport and on the snowball bifurcation for a tidally locked planet. In particular, whether the substellar region is dominated by land mass or ocean will affect evaporation and atmospheric water vapor \citep{chen2018biosignature}, which may influence the balance of latent heat-driven atmospheric vs. oceanic heat transport.

In our simulations, we kept rotation rate (and therefore orbital period) fixed as we varied stellar irradiation. This was similarly done in previous work by \citet{way2015exploring, fujii2017nir,yang2019ocean}. The advantage of this setup is that it allows us to isolate the effects of changes in radiative forcing when looking for a snowball bifurcation. In reality, different stellar irradiations (and therefore different orbital radii) correspond to different rotation rates/orbital periods. Future work could self-consistently vary rotation rate and stellar irradiation, which may impact planetary climate \citep{kumar2016inner, haqq2018demarcating,wolf2017constraints}.

 \section{Conclusions}

The main conclusion of this paper is that we do not find a snowball bifurcation for tidally locked planets using a coupled ocean-atmosphere GCM (ROCKE-3D). Habitable tidally locked planets are therefore unlikely to be found in a snowball state for a geologically significant period of time.

\section{Acknowledgments}

We thank Jun Yang and Yaoxuan Zeng for helpful discussion and comments on this work. We thank Thaddeus Komacek and Jeffrey Yang for their contributions in setting up ROCKE-3D for our study. We acknowledge support from NASA grant number NNX16AR85G, which is part of the “Habitable Worlds” program. This work was supported by the NASA Astrobiology Program Grant Number
80NSSC18K0829 and benefited from participation in the NASA Nexus for
Exoplanet Systems Science research coordination network. SLO acknowledges support from the T.C. Chamberlin Postdoctoral Fellowship in the Department of Geophysical Sciences at the University of Chicago.

\bibliography{bib}

\end{document}